%% file: 0main.tex
\def\year{2020}\relax
\documentclass[letterpaper]{article} 
\usepackage{aaai20}  
\usepackage{times}  
\usepackage{helvet} 
\usepackage{courier}  
\usepackage[hyphens]{url}  
\usepackage{graphicx}  
\usepackage{booktabs} 
\usepackage{amsmath}
\usepackage{amsfonts}
\usepackage{comment}
\usepackage{footmisc}
\usepackage{mathrsfs}
\usepackage{amssymb}
\usepackage{graphicx}
\usepackage{subfigure}
\usepackage{multirow}
\usepackage{algorithm}
\usepackage{algorithmic}
\usepackage{multicol}  
\usepackage{tikz}
\usetikzlibrary{bayesnet}
\usepackage{enumitem}
\usepackage{array}
\usepackage{float}
\usepackage{hyperref}
\begin{document}
\title{Simulating User Feedback for Reinforcement Learning Based Recommendations}
\author{Submitted for Blind Review}
\author{
	\Large \textbf{Xiangyu Zhao\textsuperscript{\rm 1}, Long Xia\textsuperscript{\rm 2}, Lixin Zou\textsuperscript{\rm 3}, Dawei Yin\textsuperscript{\rm 2}, Jiliang Tang\textsuperscript{\rm 1}}\\	
	\textsuperscript{\rm 1}Michigan State University, \textsuperscript{\rm 2}JD.com, \textsuperscript{\rm 3}Tsinghua University\
}
\maketitle 
\begin{abstract}
With the recent advances in Reinforcement Learning (RL), there have been tremendous interests in employing RL for recommender systems. However, directly training and evaluating a new RL-based recommendation algorithm needs to collect users' real-time feedback in the real system, which is time and efforts consuming and could negatively impact on users' experiences. Thus, it calls for a user simulator that can mimic real users' behaviors where we can pre-train and evaluate new recommendation algorithms. Simulating users' behaviors in a dynamic system faces immense challenges -- (i) the underlining item distribution is complex, and (ii) historical logs for each user are limited. In this paper, we develop a user simulator base on Generative Adversarial Network (GAN). To be specific, the generator captures the underlining distribution of users' historical logs and generates realistic logs that can be considered as augmentations of real logs; while the discriminator not only distinguishes real and fake logs but also predicts users' behaviors. The experimental results based on real-world e-commerce data demonstrate the effectiveness of the proposed simulator. 
\end{abstract}

\input{1introduction}
\input{2problem}
\input{3model}

\input{5experiment}
\input{6relatedwork}

\input{7conclusion}
\bibliographystyle{aaai}
\bibliography{9reference}
\end{document}

%% file: 1introduction.tex
\section{Introduction}
\label{sec:introduction}
With the recent tremendous development in Reinforcement Learning (RL), there have been increasing interests in adapting RL for recommendations. RL-based recommender systems treat the recommendation procedures as sequential interactions between users and a recommender agent (RA). They aim to automatically learn an optimal recommendation strategy (policy) that maximizes cumulative reward from users without any specific instructions. RL-based recommender systems can achieve two key advantages: (i) the recommender agent can learn their recommendation strategies based on users' real-time feedback during the user-agent interactions continuously; and (ii) the optimal strategies target at maximizing the long-term reward from users (e.g. the overall revenue of a recommendation session). Therefore, numerous efforts have been made on developing RL-based recommender systems~\cite{dulac2015deep,zhao2018deep,zheng2018drn}. 


\begin{figure}[t]
	\centering
	\includegraphics[width=66mm]{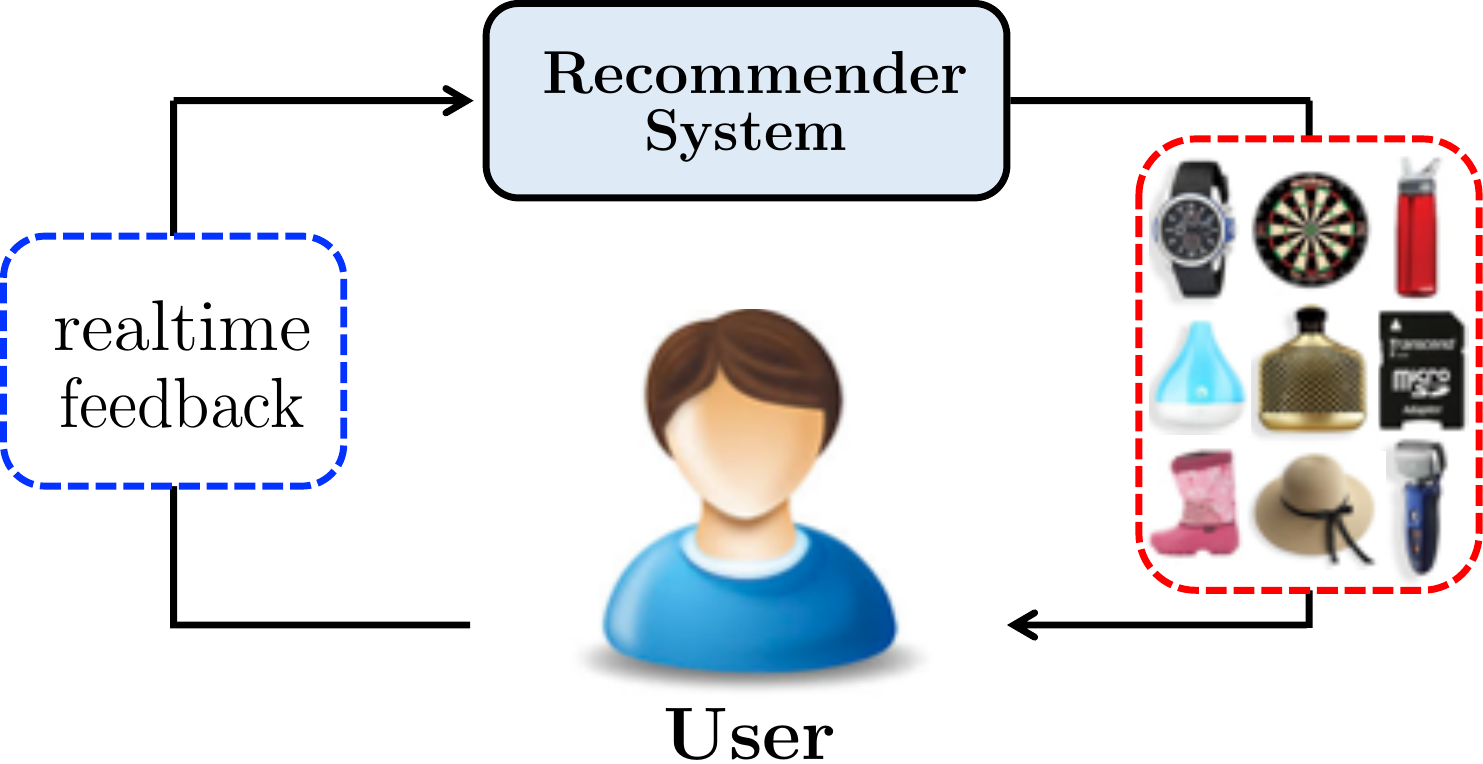}
	\caption{An example of system-user interactions.}
	\label{fig:mdp_example}
\end{figure}

RL-based recommendation algorithms are desired to be trained and evaluated based on users' real-time feedback (reward function). The most practical way is online A/B test, where a new recommendation algorithm is trained based on the feedback from real users and the performance is compared against that of the previous algorithm via randomized experiments. However, online A/B tests are inefficient and expensive: (1) online A/B tests usually take several weeks to collect sufficient data for sake of statistical sufficiency, and (2) numerous engineering efforts are required to deploy the new algorithm in the real system~\cite{yang2018unbiased,gilotte2018offline,li2015toward}. Furthermore, online A/B tests often lead to bad user experience in the initial stage when the new recommendation algorithms have not been well trained~\cite{li2012unbiased}. These reasons prevent us from quickly training and testing new RL-based recommendation algorithms. One successful solution to handle these challenges in the RL community is to first build a simulator to approximate the environment (e.g. OpenAI Gym for video games), and then use it to train and evaluate the RL algorithms~\cite{gao2014machine}. Thus, following the best practice, we aim to build a user simulator based on users' historical logs in this work, which can be utilized to pre-train and evaluate new recommendation algorithms before launching them online.

However, simulating users' behaviors (feedback) in a dynamic recommendation environment is very challenging. There are millions of items in practical recommender systems. Thus the underlining distribution of recommended item sequences are widely spanned and extremely complex in historical logs. In order to learn a robust simulator, it typically requires large-scale historical logs as training data from each user. Though massive historical logs are often available, data available to each user is rather limited. An attempt to tackle the two aforementioned challenges, we propose a simulator (RecSimu) for reinforcement learning based recommendations based on Generative Adversarial Network (GAN)~\cite{goodfellow2014generative}. We summarize our major contributions as follows:

\begin{itemize}[leftmargin=*]
	\item We introduce a principled approach to capture the underlining distribution of recommended item sequences in historical logs, and generate realistic item sequences;
	\item We propose a user behavior simulator RecSimu, which can be utilized to simulate environments to pre-train and evaluate RL based recommender systems; and  
	\item We conduct experiments based on real-world data to demonstrate the effectiveness of the proposed simulator and validate the effectiveness of its components.
\end{itemize}


%% file: 2problem.tex
\begin{figure}[t]
	\centering
	\includegraphics[width=81mm]{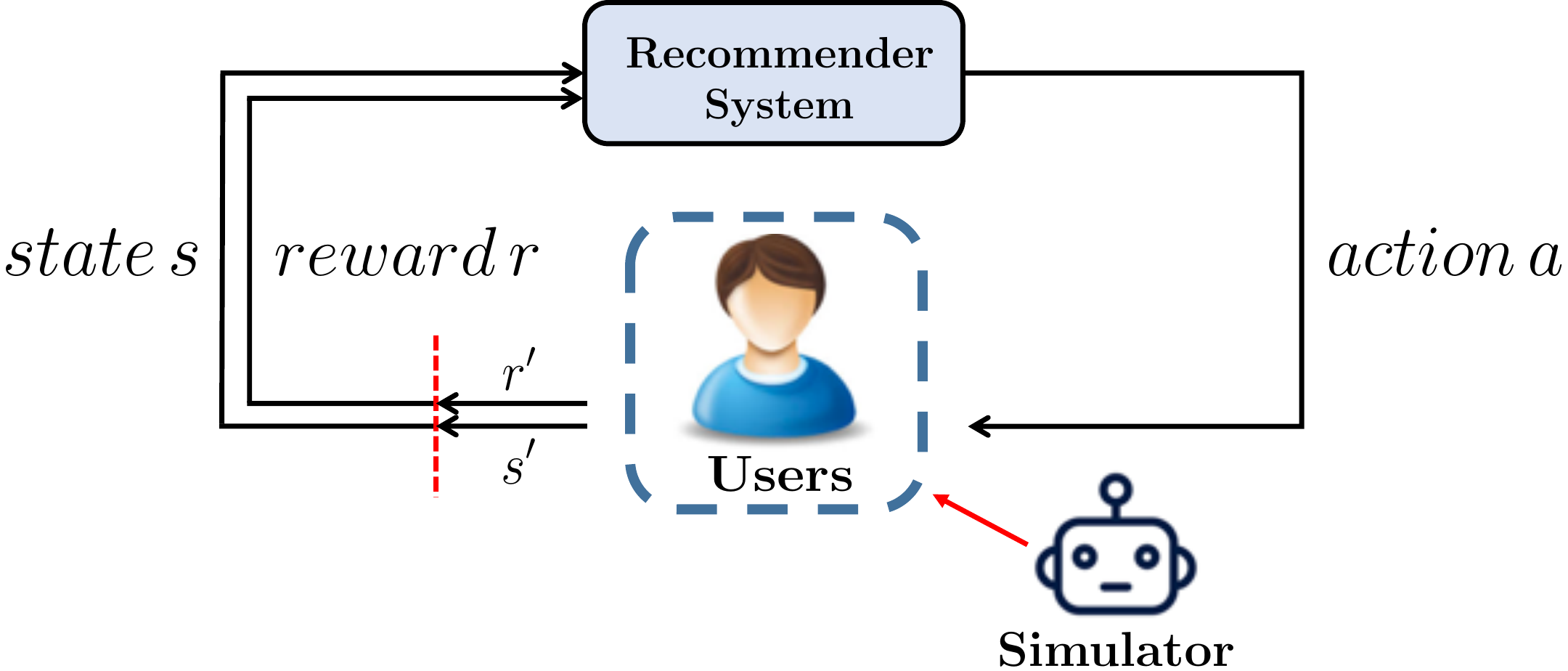}
	\caption{A common setting for RL-based RS.}
	\label{fig:interaction}
\end{figure}

\section{Problem Statement}
\label{sec:problem}

Following one common setting as shown in Figure~\ref{fig:interaction}, we first formally define reinforcement learning based recommendations~\cite{zhao2018deep,zhao2018recommendations} and then present the problem we aim to solve based on this setting. In this setting, we treat the recommendation task as sequential interactions between a recommender system (agent) and users (environment $\mathcal{E}$), and use Markov Decision Process (MDP) to model it.  It consists of a sequence of states, actions and rewards. Typically, MDP involves four elements $(\mathcal{S}, \mathcal{A}, \mathcal{P}, \mathcal{R})$, and below we introduce how to set them.  Note that there are other settings~\cite{dulac2015deep,zheng2018drn} and we leave further investigations on them as one future work.

\begin{itemize}[leftmargin=*]
	\item {\bf State space $\mathcal{S}$}: We define the state $s = \{i_{1}, \cdots, i_{N}\} \in \mathcal{S}$ as a sequence of $N$ items that a user browsed and user's corresponding feedback for each item. The items in $s$ are chronologically sorted.  
	\item {\bf Action space $\mathcal{A}$}:  An action $a \in \mathcal{A}$ from the recommender system perspective is defined as recommending a set of items to a user. Without loss of generality, we suppose that each time the recommender system suggests one item to the user, but it is straightforward to extend this setting to recommending more items.  
	\item {\bf Reward $\mathcal{R}$}: When the system takes an action $a$ based on the state $s$, the user will browse the recommended item and provide her feedback of the item. In this paper, we assume a user could skip, click and purchase the recommended items. Then the recommender system will receive a reward $r(s,a)$ solely according to the type of feedback.
	\item {\bf State transition probability $\mathcal{P}$}: State transition probability $p(s'|s,a)$ is defined as the probability that the state transits from $s$ to $s'$ when action $a$ is executed. We assume that the state transition is deterministic: we always remove the first item $i_{1}$ from $s$ and add the action $a$ at the bottom of $s$, i.e., $s' = \{i_{2}, \cdots, i_{N},a\}$.
\end{itemize}

With the aforementioned definitions and notations, in this paper, we aim to build a simulator to imitate users' feedback (behavior) on a recommended item according to user's preference learned from the user's browsing history as demonstrated in Figure~\ref{fig:interaction}. In other words, the simulator aims to mimic the reward function $r(s,a)$.  More formally, the goal of a simulator can be formally defined as follows: \textit{Given a state-action pair $(s, a)$, the goal is to find a reward function $r(s,a)$, which can accurately imitate user's behaviors.} 

%% file: 3model.tex
\begin{figure}[t]
	\centering
	\includegraphics[width=81mm]{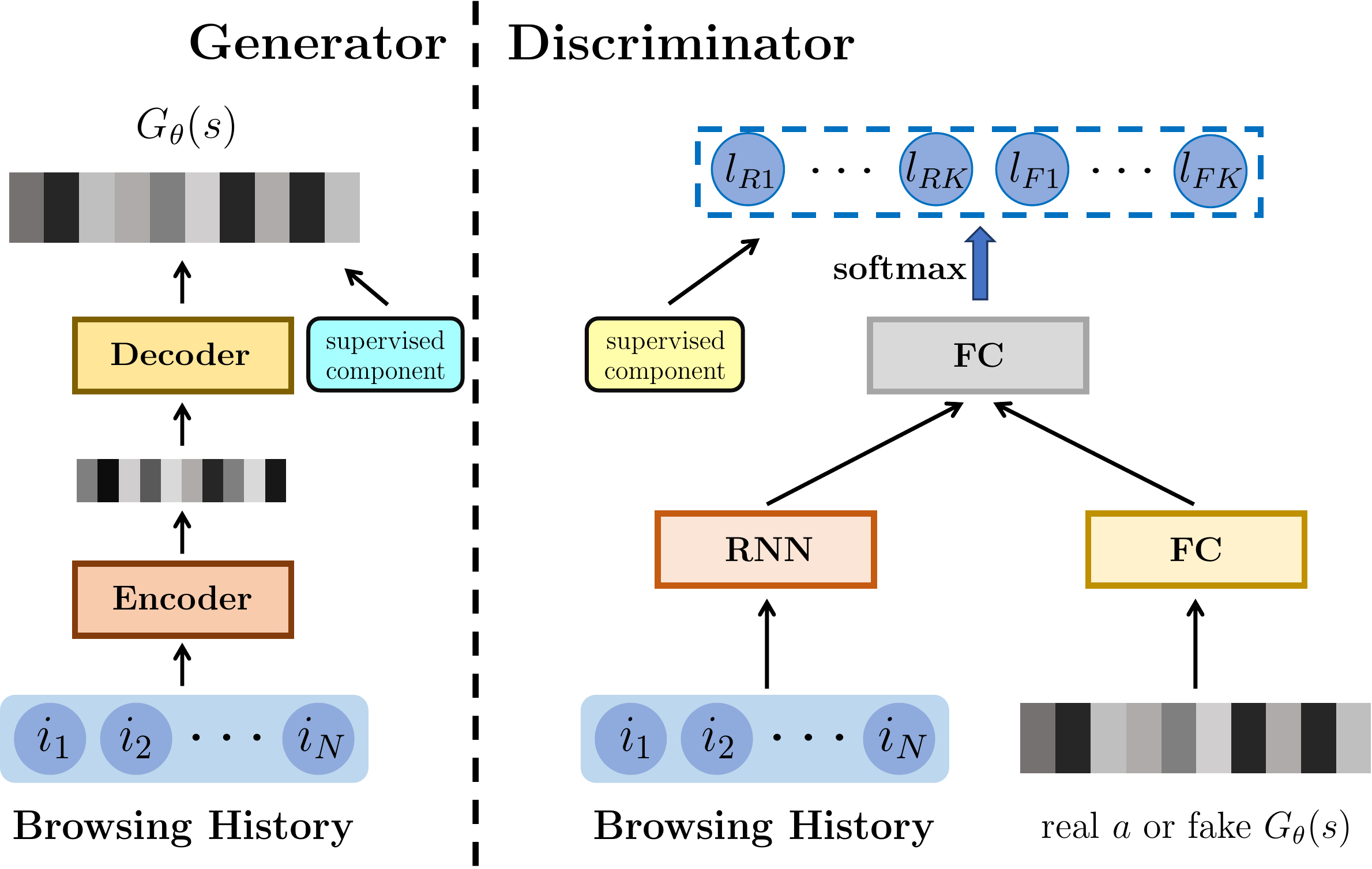}
	\caption{An overview of the proposed simulator.\label{fig:GAN}}
\end{figure}

\section{The Proposed Simulator}
\label{sec:model}
In this section, we will propose a simulator framework that imitates users' feedback (behavior) on a recommended item according to the user's current preference learned from her browsing history. As aforementioned, building a user simulator is challenging, since (1) the underlining distribution of item sequences in users' historical logs is complex, and (2) historical data for each user is usually limited.

Recent efforts have demonstrated that Generative Adversarial Network (GAN) and its variants are able to generate fake but realistic images~\cite{goodfellow2014generative}, which implies their potential in modeling complex distributions. Furthermore, the generated images can be considered as augmentations of real-world images to enlarge the data space. Driven by these advantages, we propose to build the simulator based on GAN to capture the complex distribution of users' browsing logs and generate realistic logs to enrich the training dataset. Another challenge with GAN-based simulator is that the discriminator should not only be able to distinguish real logs and generated logs, but also can predict user's feedback of a recommended item. To address these challenges, we propose a recommendation simulator as shown in Figure~\ref{fig:GAN}, where the generator with a supervised component is designed to learn the data distribution and generate indistinguishable logs, and the discriminator with a supervised component can simultaneously distinguish real/generated logs and predict user's feedback of a recommended item. In the following, we will first introduce the architectures of generator and discriminator separately, and then discuss the objective functions with the optimization.

\begin{figure}[t]
	\centering
	\includegraphics[width=95mm]{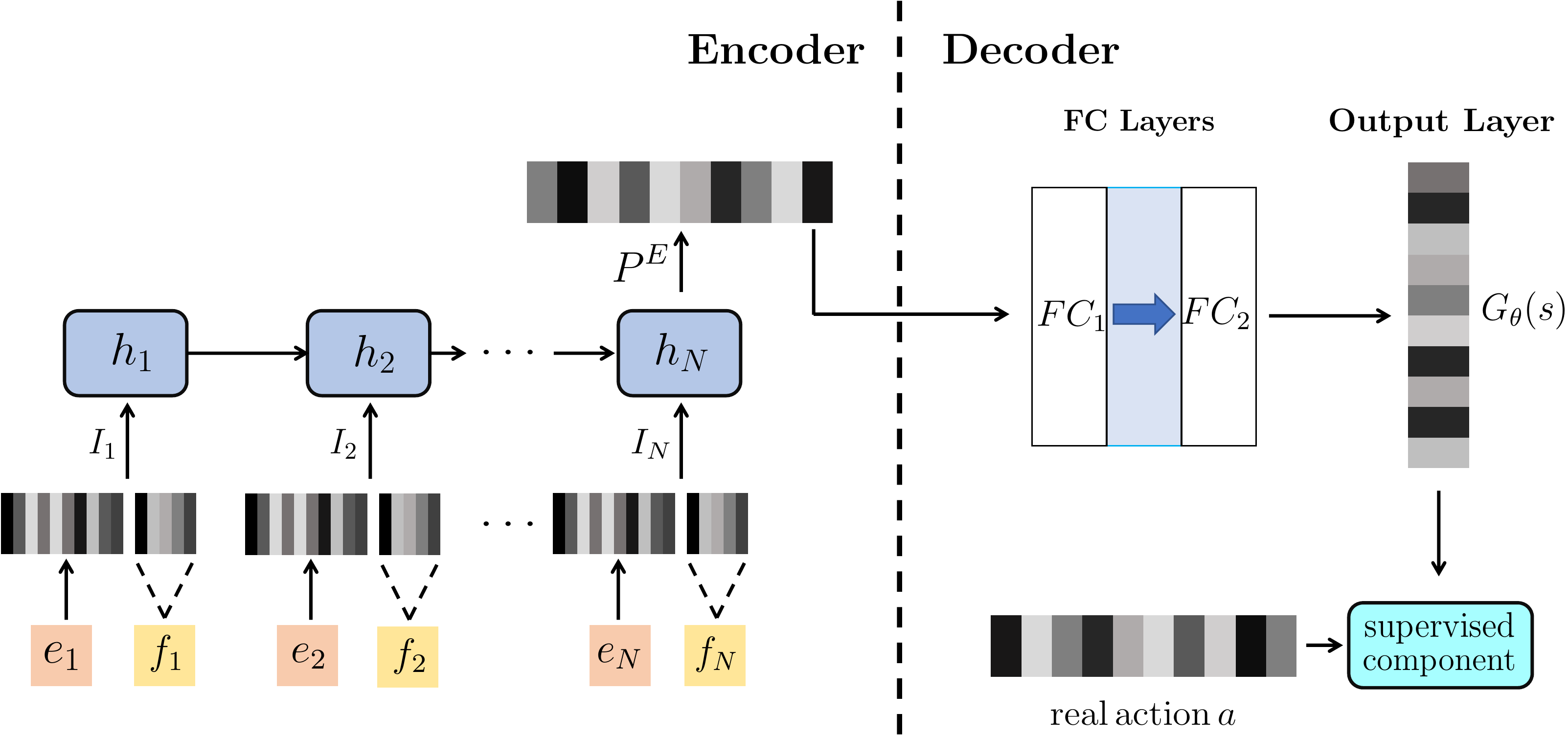}
	\caption{The generator with Encoder-Decoder architecture.\label{fig:generator}}
\end{figure}

\subsection{The Generator Architecture}
\label{sec:generator}
The goal of the generator is to learn the data distribution and then generate indistinguishable logs (action) based on users' browsing history (state), i.e., to imitate the recommendation policy of the recommender system that generates the historical logs. Figure~\ref{fig:generator} illustrates the generator with the Encoder-Decoder architecture. The Encoder component aims to learn user's preference according to the items browsed by the user and the user's feedback. The input is the state $s= \{i_{1}, \cdots, i_{N}\}$ that is observed in the historical logs, i.e., the sequence of $N$ items that a user browsed and user's corresponding feedback for each item. The output is a low-dimensional representation of user's current preference, referred as to $p^E$. Each item $i_n \in s$ involves two types of information:
\begin{equation}\label{equ:tuple}
i_n =(e_n,f_n),
\end{equation}
where $e_n$ is a low-dimensional and dense item-embedding of the recommended item~\footnote{The item-embeddings are pre-trained by an e-commerce company via word embedding~\cite{levy2014neural} based on users' historical browsing logs, where each item is considered as a word and the item sequence of a recommendation session as a sentence. The effectiveness of these item representations is demonstrated in their business like searching, recommendation and advertisement. }, and $f_n$ is a one-hot vector representation to denote user's feedback on the recommended item. The intuition of selecting these two types of information is that, we not only want to learn the information of each item in the sequence, but also want to capture user's interests (feedback) on each item. We use an embedding layer to transform $f_n$ into a low-dimensional and dense vector: $F_n = \tanh(W_F f_n + b_F)\in \mathbb{R}^{|F|}$. Note that we use ``$\tanh$'' activate function since $e_n \in (-1,+1)$. Then, we concatenate $e_n$ and $F_n$, and get a low-dimensional and dense vector $I_n \in \mathbb{R}^{|I|}$ ($|I| = |E| + |F|$) as:
\begin{equation}\label{equ:concatenate}
	\begin{aligned}
	I_n &=concat(e_n,F_n)\\		
	&=concat(e_n,\tanh(W_F f_n + b_F)).
	\end{aligned}
\end{equation}
Note that all embedding layers share the same parameters $W_F$ and $b_F$, which can reduce the amount of parameters and have better generalization. We introduce a Recurrent Neural Network (RNN) with Gated Recurrent Units (GRU) to capture the sequential patterns of items in the logs. We choose GRU rather than Long Short-Term Memory (LSTM) because of GRU's fewer parameters and simpler architecture. We consider the final hidden state of RNN as the output of Encoder component, i.e., the lower dimensional representation of user's current preference $p^E$.

The goal of the Decoder component is to predict the item that will be recommended according to the user's current preference. Therefore, the input is user's preference representation $p^E$, while the output is the item-embedding of the item that is predicted to appear at next position in the log, referred as to $G_\theta(s)$. For simplification, we leverage several fully-connected layers as the Decoder to directly transform $p^E$ to $G_\theta(s)$. Note that it is straightforward to leverage other methods to generate the next item, such as using a $softmax$ layer to compute relevance scores of all items, and selecting the item with the highest score as the next item. So far, we have delineated the architecture of the Generator, which aims to imitate the recommendation policy of the existing recommender system, and generate realistic logs to augment the historical data. In addition, we add a supervised component to encourage the generator to yield items that are close to the ground truth items, which will be discussed in Section~\ref{sec:loss}. Next, we will discuss the architecture of discriminator.

\begin{figure}[t]
	\centering
	\includegraphics[width=81mm]{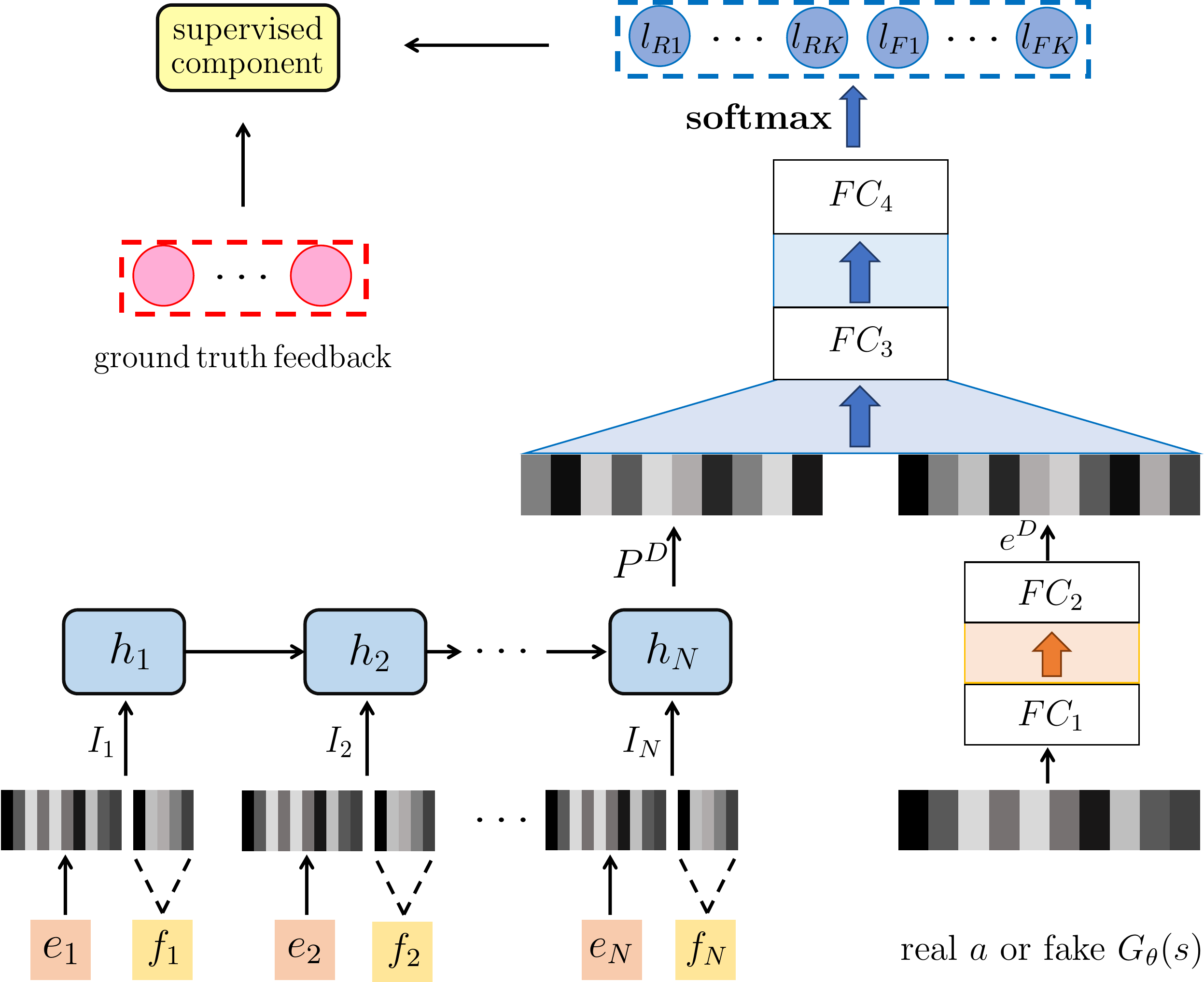}
	\caption{The discriminator architecture.\label{fig:discriminator}}
\end{figure}
\subsection{The Discriminator Architecture}
\label{sec:discriminator}
The discriminator aims to not only distinguish real historical logs and generated logs, but also predict the class of user's feedback of a recommended item according to her browsing history. Thus we consider the problem as a classification problem with $2 \times K$ classes, i.e., $K$ classes of \textit{real} feedback for the recommended items observed from historical logs, and $K$ classes of \textit{fake} feedback for the recommended items yielded by the generator. 

Figure \ref{fig:discriminator} illustrates the architecture of the discriminator. Similar with the generator, we introduce a RNN with GRU to capture user's dynamic preference. Note that the architecture is the same with the RNN in generator, but they have different parameters. The input of the RNN is the state $s= \{i_{1}, \cdots, i_{N}\}$ observed in the historical logs, where $i_n =(e_n,f_n)$, and the output is the dense representation of user's current preference, referred as to $p^D$. Meanwhile, we feed the item-embedding of the recommended item (real $a$ or fake $G_\theta(s)$) into fully-connected layers, which encode the recommended items to low-dimensional representations, referred as to $e^D$. Then we concatenate $p^D$ and $e^D$,  and feed the concatenation $(p^D,e^D)$ into fully-connected layers, whose goal is to (1) judge whether the recommended items are real or fake, and (2) predict users' feedback on these items. Therefore, the final fully-connected layer outputs a $2 \times K$ dimensional vector of logits, which represent $K$ classes of \textit{real} feedback and $K$ classes of \textit{fake} feedback respectively: 
\begin{equation}\label{equ:output}
output = [l_{R1}, \cdots, l_{RK}, l_{F1}, \cdots,  l_{FK}],
\end{equation}
where we include $K$ classes of \textit{fake} feedback in output layer rather than only one \textit{fake} class, since fine-grained distinction on fake samples can increase the power of discriminator (more details in following subsections). These logits can be transformed to class probabilities through a softmax layer, and the probability corresponding to the $j^{th}$ class is:
\begin{equation}\label{equ:softmax}
p_{model}(r=l_j|s,a)= \frac{exp(l_j)}{\sum_{k=1}^{2 \times K}exp(l_k)},
\end{equation}
where $r$ is the result of classification. The objective function is based on these class probabilities. In addition, a supervised component is introduced to enhance the user's feedback prediction and more details about this component will be discussed in Section~\ref{sec:loss}.

\subsection{The Objective Function}
\label{sec:loss}

In this subsection, we will introduce the objective functions of the proposed simulator. The discriminator has two goals: (1) distinguishing real-world historical logs and generated logs, and (2) predicting the class of user's feedback of a recommended item according to the browsing history. The first goal corresponds to an unsupervised problem just like standard GAN that distinguishes real and fake images, while the second goal is a supervised problem that minimizes the class difference between users' ground truth feedback and the predicted feedback. Therefore, the loss function $L_D$ of discriminator consists of two components. 

For the unsupervised component that distinguishes real-world historical logs and generated logs, we need calculate the probability that a state-action pair is \textit{real} or \textit{fake}. From Eq~(\ref{equ:softmax}), we know the probability that a state-action pair observed from historical logs is classified as \textit{real}, referred as to $D_\phi(s,a)$, is the summation of the probabilities of $K$ \textit{real} feedback:
\begin{equation}\label{equ:d_real}
D_\phi(s,a)= \sum_{k=1}^{K} p_{model}(r=l_k|s,a)
\end{equation}
while the probability of a \textit{fake} state-action pair where $G_\theta(s)$ action is produced by the generator, say $D_\phi(s,G_\theta(s))$, is the summation of the probabilities of $K$ \textit{real} feedback:
\begin{equation}\label{equ:d_fake}
D_\phi(s,G_\theta(s))= \sum_{k=K+1}^{2 \times K} p_{model}(r=l_k|s,G_\theta(s))
\end{equation}
Then, the unsupervised component of the loss function $L_D$ is defined as follows:
\begin{equation}\label{equ:loss_unsupervised}
\begin{aligned}
L^{unsup}_D =&-\{\mathbb{E}_{s,a\sim p_{data}}[\log D_\phi(s,a)]\\
&+\mathbb{E}_{s\sim p_{data}}[\log D_\phi(s,G_\theta(s))]\},
\end{aligned}
\end{equation}
where both $s$ and $a$ are sampled from historical logs distribution $p_{data}$ in the first term; in the second term, only $s$ is sampled from historical logs distribution $p_{data}$, while the action $G_\theta(s)$ is yielded by generator policy $G_\theta$.

The supervised component targets to predict the class of user's feedback, which is formulated as a supervised problem to minimize the class difference (i.e. the cross-entropy loss) between users' ground truth feedback and the predicted feedback. Thus it also has two terms -- the first term is the cross-entropy loss of ground truth class and predicted class for a real state-action pair sampled from real historical data distribution $p_{data}$; while the second term is the cross-entropy loss of ground truth class and predicted class for a fake state-action pair, where the action is yielded by the generator. Thus the supervised component of the loss function $L_D$ is defined as follows:
\begin{equation}\label{equ:loss_supervised}
\begin{aligned}
L^{sup}_D &=-\{\mathbb{E}_{s,a,r\sim p_{data}}[\log p_{model}(r=l_k|s,a,k\leq K)]\\		
&+\lambda \cdot \mathbb{E}_{s,r\sim p_{data}}[\log p_{model}(r=l_k|s,G_\theta(s),K<k\leq 2K\},
\end{aligned}
\end{equation}
where $\lambda$ controls the contribution of the second term. The first term is a standard cross entropy loss of a supervised problem. The intuition we introduce the second term of Eq (\ref{equ:loss_supervised}) is -- in order to tackle the data limitation challenge mentioned in Section~\ref{sec:introduction}, we consider fake state-action pairs as augmentations of real state-action pairs, then fine-grained distinction on fake state-action pairs will increase the power of discriminator, which also in turn forces the generator to output more indistinguishable actions. The overall loss function of the discriminator $L_D$ is defined as follows:
\begin{equation}\label{equ:loss_D}
L_D = L^{unsup}_D+  \alpha\cdot L^{sup}_D,
\end{equation}
where parameter $\alpha$ is introduced to control the contribution of the supervised component.

The target of the generator is to output realistic recommended items $G_\theta(s)$ that can fool the discriminator, which tackles the complex data distribution problem as mentioned in Section~\ref{sec:introduction}. To achieve this goal, we design two components for the loss function $L_G$ of the generator. The first component aims to maximize $L^{unsup}_D$ in Eq (\ref{equ:loss_unsupervised}) with respect to $G_\theta$. In other words, the first component minimizes that probabilities that fake state-action pairs are classified as \textit{fake}, thus we have: 
\begin{equation}\label{equ:loss_g1}
\begin{aligned}
L^{unsup}_G =\mathbb{E}_{s\sim p_{data}}[\log D_\phi(s,G_\theta(s))],
\end{aligned}
\end{equation}
where $s$ is sampled from real historical logs distribution $p_{data}$ and the action $G_\theta(s)$ is yielded by generator policy $G_\theta$. Inspired by a supervised version of GAN~\cite{luc2016semantic}, we introduce a supervised loss $L^{sup}_G$ as the second component of $L_G$, which is the $\ell_2$ distance between the ground truth item $a$ and the generated item $G_\theta(s)$:
\begin{equation}\label{equ:loss_g2}
\begin{aligned}
L^{sup}_G =\mathbb{E}_{s,a\sim p_{data}}  \|a -G_\theta(s)\|^2_2.
\end{aligned}
\end{equation}
where $s$ and $a$ are sampled from historical logs distribution $p_{data}$. This supervised component encourages the generator to yield items that are close to the ground truth items. The overall loss function of the discriminator $L_D$ is defined as follows:
\begin{equation}\label{equ:loss_G}
L_G = L^{unsup}_G+ \beta \cdot L^{sup}_G,
\end{equation}
where $\beta$ controls the contribution of the second component.

\begin{algorithm}[H]
	\caption{\label{alg:training} An Training Algorithm for the Proposed Simulator.}
	\raggedright
	\begin{algorithmic} [1]
		\STATE Initialize the generator $G_\theta$ and discriminator $D_\phi$ with random weights $\theta$ and $\phi$
		\STATE Sample a pre-training dataset of $s,a \sim p_{data}$
		\STATE Pre-train $G_\theta$ by minimizing $L^{sup}_G$ in Eq (\ref{equ:loss_g2})
		\STATE Generate fake-actions $G_\theta(s)\sim G_\theta$ for training $D_\phi$
		\STATE Pre-train $D_\phi$ by minimizing $L^{sup}_D$ in Eq (\ref{equ:loss_supervised})
		\REPEAT
		\FOR{d-steps}
		\STATE Sample minibatch of $s,a \sim p_{data}$
		\STATE Use current $G_\theta$ to generate minibatch of $G_\theta(s)\sim G_\theta$
		\STATE Update the $D_\phi$ by minimizing $L_D$ in Eq (\ref{equ:loss_D})
		\ENDFOR
		\FOR{g-steps}
		\STATE Sample minibatch of $s,a \sim p_{data}$
		\STATE Update the $G_\theta$ by minimizing $L_G$ in Eq (\ref{equ:loss_G})
		\ENDFOR
		\UNTIL{simulator converges}
	\end{algorithmic}
\end{algorithm}

We present our simulator training algorithm in details shown in Algorithm~\ref{alg:training}. At the beginning of the training stage, we use standard supervised methods to pre-train the generator (line 3) and discriminator (line 5). After the pre-training stage, the discriminator (lines 7-11) and generator (lines 12-15) are trained alternatively. For training the discriminator, state $s$ and real action $a$ are sampled from real historical logs, while fake actions $G_\theta(s)$ are generated through the generator. To keep balance in each d-step, we generate fake actions $G_\theta(s)$ with the same number of real actions $a$.


%% file: 5experiment.tex
\section{Experiments}
\label{sec:experiment}
In this section, we conduct extensive experiments to evaluate the effectiveness of the proposed simulator with a real-world dataset from an e-commerce site. We mainly focus on two questions: (1) how the proposed simulator performs compared to the state-of-the-art baselines for predicting user behaviors (discriminator); and (2) how the generator performs compared with representative recommender algorithms. We first introduce experimental settings. Then we seek answers to the above two questions. Finally, we study the impact of important parameters on the performance of the proposed framework. 

\subsection{Experimental Settings}
\label{sec:experimental_settings}
We evaluate our method on a dataset of July 2018 from a real e-commerce company. We randomly collect 272,250 recommendation sessions, and each session is a sequence of item-feedback pairs. After filtering out items that appear less than 5 times, there remain 1,355,255 items. For each session, we use first $N$ items and corresponding feedback as the initial state, the $N+1^{th}$ item as the first action, then we could collect a sequence of (state,action,reward) tuples following the MDP defined in Section~\ref{sec:problem}. We collect the last (state,action,reward) tuples from all sessions as the test set, while using the other tuples as the training set.   

In this paper, we leverage $N = 20$ items that a user browsed and user's corresponding feedback for each item as state $s$. The dimension of the item-embedding $e_n$ is $|E|=20$, and the dimension of action representation $F_n$ is $|F|=10$ ($f_n$ is a 2-dimensional one-hot vector: $f_n = [1,0]$ when feedback is negative, while $f_n = [0,1]$ when feedback is positive). The output of discriminator is a $4\,(K=2)$ dimensional vector of logits, and each logit represents \textit{real-positive}, \textit{real-negative}, \textit{fake-positive} and \textit{fake-negative} respectively:
\begin{equation}
output = [l_{rp}, l_{rn}, l_{fp},  l_{fn}],
\end{equation}
where \textit{real} denotes that the recommended item is observed from historical logs; \textit{fake} denotes that the recommended item is yielded by the generator; \textit{positive} denotes that a user clicks/purchases the recommended item; and \textit{negative} denotes that a user skips the recommended item. Note that though we only simulate two types of behaviors of users (i.e., positive and negative), it is straightforward to extend the simulators with more types of behaviors. AdamOptimizer is applied in optimization, and the learning rate for both Generator and Discriminator is 0.001, and batch-size is 500. The hidden size of RNN is 128. For the parameters of the proposed framework such as $\alpha$, $\beta$ and $\lambda$, we select them via cross-validation. Correspondingly, we also do parameter-tuning for baselines for a fair comparison. We will discuss more details about parameter selection for the proposed simulator in the following subsections. 

In the test stage, given a state-action pair, the simulator will predict the classes of user's feedback for the action (recommended item), and then compare the prediction with ground truth feedback observed from the historical log. For this classification task, we select the commonly used \textit{F1-score} as the metric, which is a measure that combines precision and recall, namely the harmonic mean of precision and recall.  Moreover, we leverage $p_{model}(r=l_{rp}|s,a)$ (i.e. the probability that user will provide positive feedback to a real recommended item) as the score, and use \textit{AUC} (Area under the ROC Curve) as the metric to evaluate the performance.

\subsection{Comparison of the Overall Performance}
To answer the first question, we compare the proposed simulator (discriminator) with the following state-of-the-art baseline methods:
\begin{itemize}[leftmargin=*]
	\item \textbf{Random}: This baseline randomly assigns each recommended item a $score \in [0,1]$, and uses $0.5$ as the threshold value to classify items as positive or negative; this $score$ is also used to calculate AUC. 
	\item \textbf{LR}: Logistic Regression~\cite{menard2002applied} uses a logistic function to model a binary dependent variable through minimizing the loss $\mathbb{E} \frac{1}{2} (h_\theta(x)-y)^2$, where $h_\theta(x)=\frac{1}{1+e^{-w^Tx}}$; we concatenate all $i_n =(e_n,f_n)$ as the feature vector for the $i$-th item, and set ground truth $y=1$ if feedback is positive, otherwise $y=0$.
	\item \textbf{GRU}: This baseline utilizes an RNN with GRU to predict the class of user's feedback to a recommended item. The input of each unit is $i_n =(e_n,f_n)$, and the output of RNN is the representation of user's preference, say $u$, then we concatenate $u$ with the embedding of a recommended item, and leverage a $softmax$ layer to predict the class of user's feedback to this item.
	\item \textbf{GAN}: This baseline is based on Generative Adversarial Network~\cite{goodfellow2014generative}, where the generator takes state-action pairs (the browsing histories and the recommended items) and outputs user's feedback (reward) to the items, while the discriminator takes (state, action, reward) tuples and distinguishes between real tuples (whose rewards are observed from historical logs) and fake ones.	Note that we also use an RNN with GRU to capture user's sequential preference.
	\item \textbf{GAN-s}: This baseline is a supervised version of GAN~\cite{luc2016semantic}, where the setting is similar with the above GAN baseline, while a supervise component is added on the output of the generator, which minimizes the difference between real feedback and predicted feedback.
\end{itemize}

\begin{figure}[t]
	\centering
	\includegraphics[width=81mm]{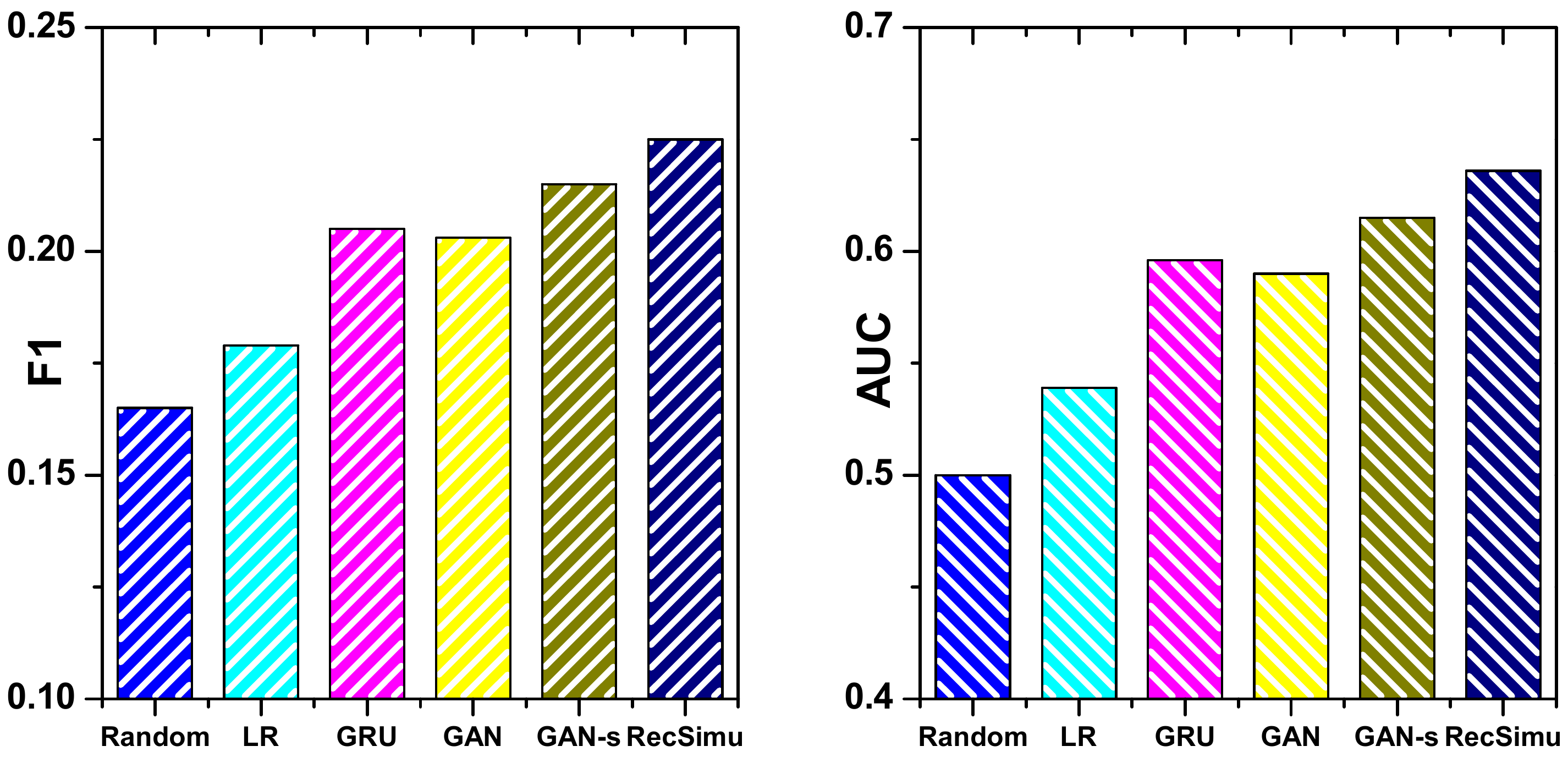}
	\caption{The results of overall performance comparison.}
	\label{fig:overall}
\end{figure}

The results are shown in Figure~\ref{fig:overall}. We make the following observations:
\begin{enumerate}[leftmargin=*]
	\item LR achieves worse performance than GRU, since LR neglects the temporal sequence within users' browsing history, while GRU can capture the temporal patterns within the item sequences and users' feedback for each item. This result demonstrates that it is important to capture the sequential patterns of users' browsing history when learning user's dynamic preference.
	\item GAN-s performs better than GRU and GAN, since GAN-s benefits from not only the advantages of the GAN framework (the unsupervised component), but also the advantages of the supervised component that directly minimizes the cross-entropy between the ground truth feedback and the predicted feedback.
	\item RecSimu outperforms GAN-s because the generator imitates the recommendation policy that generates the historical logs, and the generated logs can be considered as augmentations of real logs, which solves the data limitation challenge; while the discriminator can distinguish real and generated logs (unsupervised component), and simultaneously predict user's feedback of a recommended item (supervised component). In other words, RecSimu takes advantage of both the unsupervised and supervised components. The contributions of model components of RecSimu will be studied in the following subsection.
\end{enumerate}

To sum up, the proposed framework outperforms the state-of-the-art baselines, which validates its effectiveness in simulating users' behaviors in recommendation tasks.

\begin{table}[H]
	\centering
	\caption{Generator effectiveness results.}
	\label{table:result2}
	\begin{tabular}{ccccc}
		\hline\hline
		Recommender& MAP & diff. & NDCG@40 & diff.\\\hline\hline
		Logs& 0.097 &-&0.207&-\\\hline
		FM&0.075&22.6\%&0.161&22.2\%\\
		W\&D&0.082&15.5\%&0.176&15.0\%\\
		GRU4Rec&0.089&8.25\%&0.189&8.70\%\\
		RecSimu&\textbf{0.092}&\textbf{5.16\%}&\textbf{0.196}&\textbf{5.31\%}\\\hline\hline
	\end{tabular}
\vspace{-3mm}
\end{table}

\subsection{Effectiveness of Generator}
\label{sec:ev_generator}
Our proposed generator aims to generate indistinguishable logs (action) based on users' browsing history (state). In other words, it mimics the recommendation policy of the recommender system that generates the historical logs. To answer the second question, we train several representative recommender algorithms based on the historical logs we use in this paper and compare the performance difference with the historical logs. To evaluate the performance of the recommender algorithms, we select \textbf{MAP} and \textbf{NDCG}~\cite{jarvelin2002cumulated} as the metrics. We compare the generator of the proposed generator with the following representative recommender methods:
\begin{itemize}
	\item \textbf{FM}: Factorization Machines \cite{rendle2010factorization} combine the advantages of SVMs with factorization models. Compared with matrix factorization, higher order interactions can be modeled using the dimensionality parameter.
	\item \textbf{W\&D} \cite{cheng2016wide}: This baseline is a wide \& deep model for jointly training feed-forward neural networks with embeddings and linear model with feature transformations for generic recommender systems. 
	\item \textbf{GRU4Rec} \cite{hidasi2015session}: GRU4Rec utilizes RNN with GRU units to predict what user will click/order next based on the clicking/ordering histories. 
\end{itemize}

The results are shown in Table~\ref{table:result2}. Compared with baselines, it can be observed that the generator of RecSimu could achieve the most similar performance with the historical logs. This result validates that the competition between the generator and discriminator can enhance the generator's ability to capture the complex item distribution in historical logs over supervised recommender algorithms.

\section{Component Anslysis}
To study how the components in the generator and discriminator contribute to the performance, we systematically eliminate the corresponding components of the simulator by defining following variants of RecSimu:
\begin{itemize}[leftmargin=*]
	\item \textbf{RecSimu-1}: This variant is a simplified version of the simulator who has the same architecture except that the output of the discriminator is a 3-dimensional vector $output = [l_{rp}, l_{rn}, l_{f}]$, where each logit represents \textit{real-positive}, \textit{real-negative} and \textit{fake} respectively, i.e., it will not distinguish the generated positive and negative items.
	\item \textbf{RecSimu-2}: In this variant, we evaluate the contribution of the supervised component $L^{sup}_G$, so we eliminate the impact of $L^{sup}_G$ by setting $\beta=0$.
	\item \textbf{RecSimu-3}: This variant is to evaluate the effectiveness of the competition between generator and discriminator, hence, we remove $L^{unsup}_G$ and $L^{unsup}_D$ from loss function.
\end{itemize}

\begin{figure}[t]
	\centering
	\includegraphics[width=81mm]{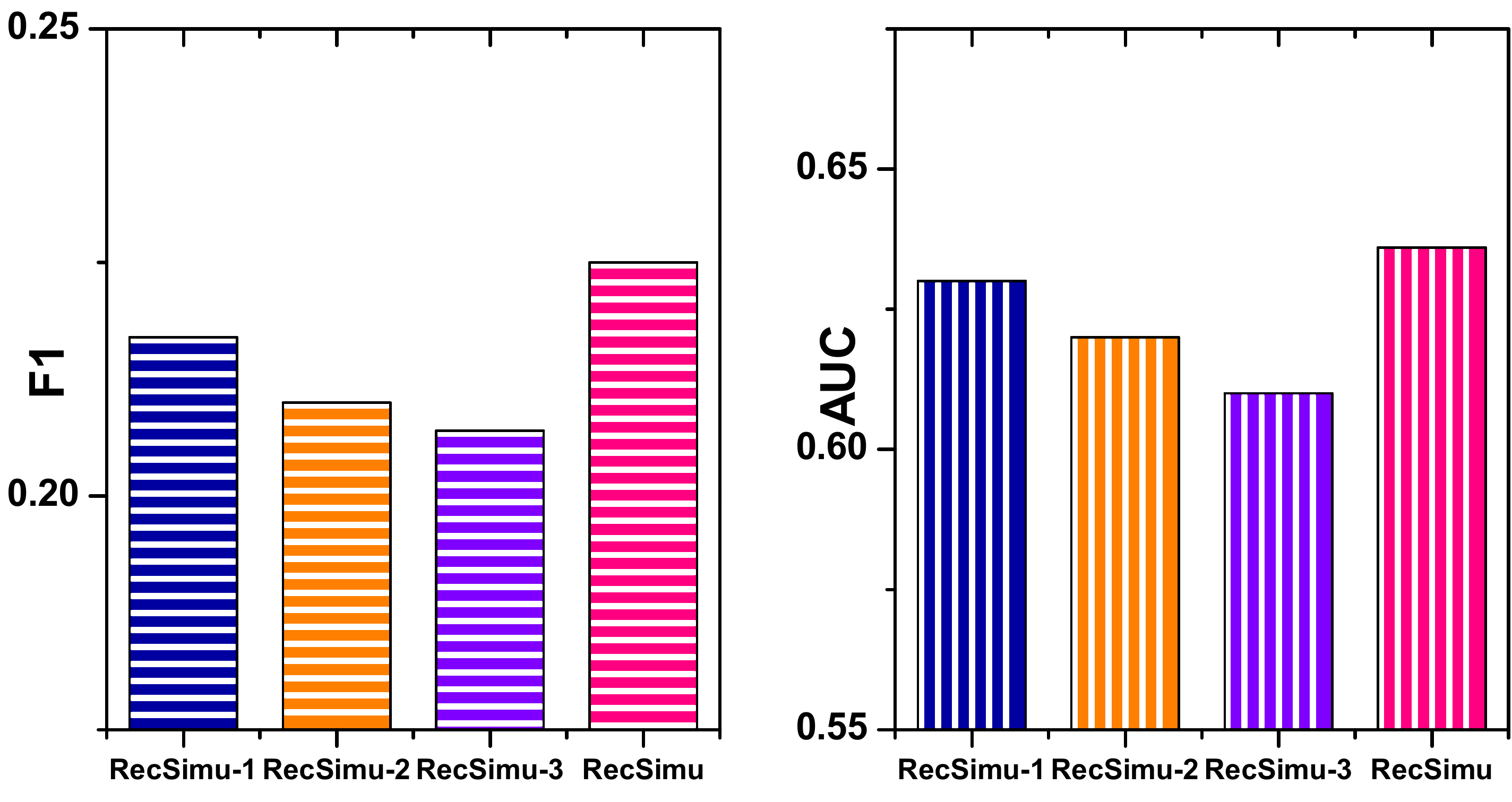}
	\caption{The results of component analysis.}
	\label{fig:component}
\end{figure}

The results are shown in Figure~\ref{fig:component}. It can be observed:

\begin{enumerate}[leftmargin=*]
	\item RecSimu performs better than RecSimu-1, which demonstrates that distinguishing the generated positive and negative items can enhance the performance. This also validates that the generated data from the generator can be considered as augmentations of real-world data, which resolves the data limitation challenge.
	\item RecSimu-2 performs worse than RecSimu, which suggests that the supervised component is helpful for the generator to produce more indistinguishable items.
	\item RecSimu-3 first trains a generator, then uses real data and generated data to train the discriminator; while RecSimu updates the generator and discriminator iteratively. RecSimu outperforms RecSimu-3, which indicates that the competition between the generator and discriminator can enhance the power of both the generator (to capture complex data distribution) and the discriminator (to classify real and fake samples).
\end{enumerate}

\begin{figure}[t]
	\centering
	\includegraphics[width=81mm]{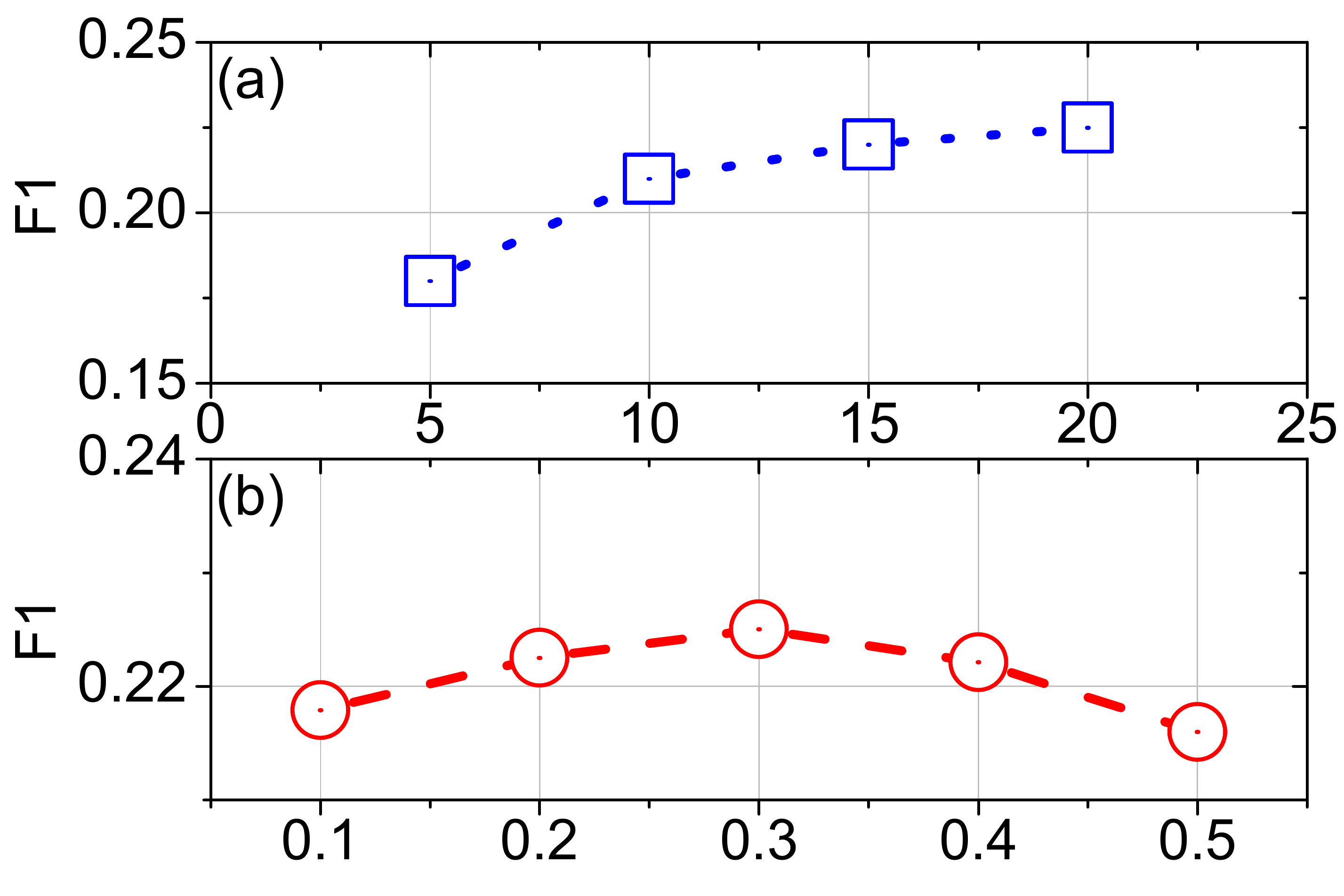}
	\caption{The results of parametric sensitivity analysis.}
	\label{fig:parameter}
\end{figure}

\subsection{Parametric Sensitivity Analysis}
\label{sec:ev_parametric}
Our method has two key parameters, i.e., (1) $N$ that controls the length of state, and (2) $\lambda$ that controls the contribution of the second term in Eq (\ref{equ:loss_supervised}), which classifies the generated items into positive or negative class. To study the impact of these parameters, we investigate how the proposed framework RecSimu works with the changes of one parameter, while fixing other parameters. The results are shown in Figure~\ref{fig:parameter}. We have following observations:

\begin{enumerate}[leftmargin=*]
\item Figure \ref{fig:parameter} (a) demonstrates the parameter sensitivity of $N$. We find that with the increase of $N$, the performance improves. To be specific, the performance improves significantly first and then becomes relatively stable. This result indicates that introducing longer browsing history can enhance the performance.
\item Figure \ref{fig:parameter} (b) shows the sensitivity of $\lambda$. The performance for the simulator achieves the peak when $\lambda = 0.3$. In other words, the second term in Eq (\ref{equ:loss_supervised}) indeed improves the performance of the simulator; however, the performance mainly depends on the first term in Eq (\ref{equ:loss_supervised}), which classifies the real items into positive and negative classes.
\end{enumerate}

%% file: 6relatedwork.tex
\section{Related Work}
\label{sec:related_work}
In this section, we briefly review works related to our study. 
In general, the related work can be mainly grouped into the following categories.

The third category related to this paper is reinforcement learning based recommender systems, which typically consider the recommendation task as a Markov Decision Process (MDP), and model the recommendation procedure as sequential interactions between users and recommender system~\cite{zhao2019deep,zhao2018reinforcement}. Practical recommender systems are always with millions of items (discrete actions) to recommend~\cite{zhao2016exploring,guo2016cosolorec}. Thus, most RL-based models will become inefficient since they are not able to handle such a large discrete action space. A Deep Deterministic Policy Gradient (DDPG) algorithm is introduced to mitigate the large action space issue in practical RL-based recommender systems~\cite{dulac2015deep}. To avoid the inconsistency of DDPG and improve recommendation performance, a tree-structured policy gradient is proposed in~\cite{chen2018large}. Biclustering technique is also introduced to model recommender systems as grid-world games so as to reduce the state/action space~\cite{choi2018reinforcement}. To solve the unstable reward distribution problem in dynamic recommendation environments, approximate regretted reward technique is proposed with Double DQN to obtain a reference baseline from individual customer sample~\cite{chen2018stabilizing}. Users' positive and negative feedback, i.e., purchase/click and skip behaviors, are jointly considered in one framework to boost recommendations, since both types of feedback can represent part of users' preference~\cite{zhao2018recommendations}. Architecture aspect and formulation aspect improvement are introduced to capture both positive and negative feedback in a unified RL framework. A page-wise recommendation framework is proposed to jointly recommend a page of items and display them within a 2-D page~\cite{zhao2017deep,zhao2018deep}. CNN technique is introduced to capture the item display patterns and users' feedback of each item in the page. A multi-agent model-based reinforcement learning framework (DeepChain) is proposed for the whole-chain recommendation problem~\cite{zhao2019model}, which is able to collaboratively train multiple recommendation agents for different scenarios by a model-based optimization algorithm. A user simulator RecSimu base on Generative Adversarial Network (GAN) framework is presented for RL-based recommender systems~\cite{zhao2019toward}, which models real users' behaviors from users' historical logs, and tackle the two challenges: (i) the recommended item distribution is complex within users' historical logs,  and (ii) labeled training data from each user is limited. In the news feed scenario, a DQN based framework is proposed to handle the challenges of conventional models, i.e., (1) only modeling current reward like CTR, (2) not considering click/skip labels, and (3) feeding similar news to users~\cite{zheng2018drn}. An RL framework for explainable recommendation is proposed in~\cite{wang2018reinforcement}, which can explain any recommendation model and can flexibly control the explanation quality based on the application scenario. A policy gradient-based top-K recommender system for YouTube is developed in~\cite{chen2018top}, which addresses biases in logged data through incorporating a learned logging policy and a novel top-K off-policy correction. Other applications includes sellers' impression allocation~\cite{cai2018reinforcement}, fraudulent behavior detection~\cite{cai2018reinforcement1}, and user state representation~\cite{liu2018deep}.

The second category related to this paper is behavior simulation.
One of the most effective approaches is Learning from Demonstration (LfD), which estimates implicit reward function from expert's behavior state to action mappings. Successful LfD applications include autonomous helicopter maneuvers~\cite{ross2013learning}, self-driving car~\cite{bojarski2016end}, playing table tennis~\cite{calinon2010learning}, object manipulation~\cite{pastor2009learning} and making coffee~\cite{sung2018robobarista}. For example, Ross et al.~\cite{ross2013learning} develop a method that autonomously navigates a small helicopter at low altitude in a natural forest environment. 
Bojarski et al.~\cite{bojarski2016end} train a CNN to directly map the raw pixels of a single front-facing camera to the steering commands. 
Calinon et al.~\cite{calinon2010learning} propose a probabilistic method to train robust models of human motion by imitating, e.g., playing table tennis. 
Pastor et al.~\cite{pastor2009learning} present a general method to learn robot motor skills from human demonstrations. 
Sung et al.~\cite{sung2018robobarista} proposed a manipulation planning approach according to the assumption that many household items share similar operational components. 

%% file: 7conclusion.tex
\section{Conclusion}
\label{sec:conclusion}
In this paper, we propose a novel user simulator RecSimu base on Generative Adversarial Network (GAN) framework, which models real users' behaviors from users' historical logs, and tackle the two challenges: (i) the recommended item distribution is complex within users' historical logs,  and (ii) labeled training data from each user is limited. The GAN-based user simulator can naturally resolve these two challenges and can be used to pre-train and evaluate new recommendation algorithms before launching them online. To be specific, the generator captures the underlining item distribution of users' historical logs and generates indistinguishable fake logs that can be used as augmentations of real logs; and the discriminator is able to predict users' feedback of a recommended item based on users' browsing logs, which takes advantage of both supervised and unsupervised learning techniques. In order to validate the effectiveness of the proposed user simulator, we conduct extensive experiments based on real-world e-commerce dataset. The results show that the proposed user simulator can improve the user behavior prediction performance in recommendation task with significant margin over representative baselines.

There are several interesting research directions. First, for the sake of generalization, in this paper, we do not consider the dependency between consecutive actions, in other words, we split one recommendation session to multiple independent state-action pairs. Some recent techniques of imitation learning, such as Inverse Reinforcement Learning and Generative Adversarial Imitation Learning, consider a sequence of state-action pairs as a whole trajectory and the prior actions could influence the posterior actions. We will introduce this idea as one future work. Second, positive (click/purchase) and negative (skip) feedback is extremely unbalanced in users' historical logs, which makes it even harder to collect sufficient positive feedback data. In this paper, we leverage traditional up-sampling techniques to generate more training data of positive feedback. In the future, we consider leverage the GAN framework to automatically generate more data of positive feedback. Finally, users skip items for many reasons, such as (1) users indeed don't like the item, (2) users do not look the item in detail and skip it by mistake, (3) there exists a better item in the nearby position, etc. These reasons result in predicting skip behavior even harder. Thus, we will introduce explainable recommendation techniques to identify the reasons why users skip items.